\documentclass[aps,preprint,unsortedaddress,floats]{revtex4}
\usepackage{amsmath,amsfonts,graphicx}
\begin{document}
\bibliographystyle{revtex}
\draft

\topmargin -1cm \oddsidemargin -1cm \evensidemargin -1cm

\title{SOLITON-LIKE SOLUTIONS OF THE GRAD-SHAFRANOV EQUATION}
\author{Giovanni Lapenta}
\email{lapenta@lanl.gov} \affiliation{Plasma Theory Group,
Theoretical Division, Los Alamos National Laboratory, Mail Stop:
K717, Los Alamos, NM87545, USA. }

\date{\today}

\begin{abstract}
A new class of soliton-like solutions is derived for the
Grad-Shafranov (GS) equations. A mathematical analogy between the
GS equation for MHD equilibria and the cubic Schr\"odinger (CS)
equation for non-linear wave propagation forms  the basis to
derive the new class of solutions. The soliton-like solutions are
considered for their possible relevance to astrophysics and solar
physics problems. We discuss how a soliton-like solution can  be
generated by a repetitive process of magnetic arcade stretching
and plasmoid formation induced by the differential rotation of the
solar photosphere or of an accretion disk.
\end{abstract}

\pacs{PACS numbers: }

\maketitle

We present a new class of soliton-like solutions of the
Grad-Shafranov equation. The Grad-Shafranov (GS) equation governs
the equilibrium conditions of a magnetic field embedded in a
plasma.

Countless applications rely on the GS equation to determine the
equilibrium steady state condition~\cite{biskamp}. The equilibrium
field in magnetic fusion devices is computed with the GS equation.
In the present letter, the attention focuses more on problems
where the GS equation can be of importance for space and
astrophysical systems.

In the solar corona, field lines emerging from the photosphere
form arcades whose evolution is crucial in  solar flares and
coronal mass ejections~\cite{priest}. Similarly, magnetic fields
tied to an accretion disk rotating around a massive object (such
as a star or or a black hole) can form magnetic
arcades~\cite{lyndenbell}. In both cases, photospheric motions or
the differential rotation of the accretion disk cause a shear of
the footpoints of the magnetic arcade. The shear motion, in turn,
injects toroidal magnetic flux and causes
 the expansion of the arcade away from the photosphere or from
 the accretion disk.

In this scenario, the role of the GS equation has been traditionally tied
to the understanding of the expansion of the magnetic arcade. A
series of papers have investigated the expansion phase using
self-similar solutions of the GS equation, proving that the field
expands away from a differentially rotating surface at a
characteristic angle of 60 degrees~\cite{lbb}. However, the
applicability of the GS equation to this scenario is intrinsically
limited by two factors. First, the GS neglects inertia effects
that can be important in determining the expansion of the arcade.
Secondly, the presence of dissipation processes (such as anomalous
resistivity or kinetic effects) can lead to reconnection processes
that alter the topology of the expanding arcade. Several papers
have investigated this process proving that reconnection can lead
to the formation and ejection of a plasmoid, i.e. a blob of plasma
encircled by field lines at least partially detached from the
original footpoints. In the solar case such plasmoid can be
interpreted as a coronal mass ejection.

Recently it has been pointed out that the process of arcade
expansion and plasmoid ejection can repeat itself in time at the
same location~\cite{choe}. There is observational evidence to
support this claim. In the solar case, it has been observed that
flares can repeat at the same location on a cycle of several hours
to few days~\cite{choe}. In the case of accretion disks, the repetitive
ejections of plasma "bullets" has been observed for
SS433~\cite{chakrabarti} and jets emitted from accretion disks
often present a knotty structure that suggests the presence of
multiple islands~\cite{knot}.

In the present letter, we discuss a new model of such ripetitive
series of plasmoids. The model ineterprets such observational
occurencies in the mathematical framework of soliton theory. A jet
of plasma composed by a series of plasmoids is represented
mathematically as a soliton being propagated from the source (the
photosphere in the solar case and the accretion disk for the
astrophysical case).

Below, we show that a mathematical analogy exists between the GS
equation and the cubic Schr\"odinger (CS) equation typical of the
soliton theory~\cite{lamb}. Such partial analogy is used to find a
class of exact solutions of the GS equation that have the same
mathematical structure of the soliton solutions of the CS
equation. Finally we discuss the likelihood of such solutions
existing in practice.

{\it Grad-Shafranov Equation}: Magnetic equilibria are computed
using the GS equation. The GS equation is derived
straight-forwardly  from the momentum equation assuming
equilibrium conditions and neglecting inertia:
\begin{equation}
{\bf J} \times {\bf B} = \nabla p \label{momentum}
\end{equation}

Assuming a Cartesian geometry with axis $(x,y,z)$, we assume all
quantities to depend only on $(x,y)$ and assume the third
dimension to be a symmetry direction. In most of the applications,
axisymmetric coordinates would be a more appropriate choice, but
not within the scope of the present work. The use of the
axisymmeric coordinates would obscure the mathematical equivalence
that we want to prove between the GS equation and the CS equation.
The extension of the derivations below to axisymmetric
coordinates will be the topic of a future
paper.

With the geometry chosen above, a solution of eq.~(\ref{momentum})
can be found in the form~\cite{finn}:
\begin{equation}
{\bf B} = \hat{\bf z} \times \nabla \Psi + B_z \hat{\bf z}
\label{field}
\end{equation}
where $\hat{\bf z}$ is the unit vector in the ignorable direction
$z$. From the magnetic field given in eq.~(\ref{field}), the
current follows immediately:
\begin{equation}
{\bf J} =\nabla \times {\bf B} = \nabla B_z \times \hat{\bf z}  +
\nabla^2 \Psi \hat{\bf z} \label{current}
\end{equation}

Substituting the expression for the current and the field, the
balance eq.~(\ref{momentum}) assumes a new form:
\begin{equation}
\left(\nabla B_z \times \hat{\bf z} \right) \cdot \nabla \Psi
\hat{\bf z} = \nabla p + \nabla^2 \Psi \nabla \Psi + B_z \nabla
B_z \label{GScomplete}
\end{equation}

From this rather involved equation, the GS equation follows,
noting that the left hand side is directed along the $z$ direction
and the right hand side is directed on the $(x,y)$ plane. It
follows that the two sides of the equations  must both be zero for
the complete equation to be satisfied:
\begin{equation}
\left \{\begin{array}{l} \displaystyle \left(\nabla \Psi \times
\nabla B_z \right) \cdot \hat{\bf z} = 0 \\ \\ \displaystyle
 \nabla p + \nabla^2
\Psi \nabla \Psi + B_z \nabla B_z =0   \end{array} \right.
\label{GSdue}
\end{equation}
where the left hand side of eq.~(\ref{GScomplete}) has been
manipulated into a more convenient form.

The two equations must be both satisfied to obtain a complete
equilibrium. Note that the first equation simply states that the
gradients of $B_z$ and of $\Psi$ must be parallel. This
requirement is typically enforced by requiring that $B_z$ is
itself a function of $\Psi$ (the same being also true for $p$):
\begin{equation}
\left \{\begin{array}{l} \displaystyle p=g(\Psi) \,,\; B_z=f(\Psi) \\ \\
\displaystyle
 \nabla^2
\Psi   = -g^\prime - ff^\prime  \end{array} \right. \label{GS}
\end{equation}
where the prime represent derivative with respect to $\Psi$.
Equation (\ref{GS})  is the classic form of the GS equation.

Typically it is convenient to work on the classic form (\ref{GS})
of the GS equation obtained above, but we will show that it has a
limitation: it does not allow to consider the extension of $\Psi$
in the complex plane. Often in mathematical physics, it can be
advantageous to extend  an equation in the complex plane to obtain
its solutions, returning to the real axis after the solution is
obtained. To obtain the class of soliton-like solutions derived
below, that is indeed the procedure to follow. To that end, the
original form (\ref{GSdue}) of the GS equations is to be
preferred, as the independent variables $(x,y)$ remain real even
when the dependent variable $\Psi$ is extended in the complex
plane. In the classic form of the GS equation, the extension of
$\Psi$ to the complex plane would entice the consideration of
complex functions of complex variable. The use of the original
form (\ref{GSdue}) requires only the consideration of complex
functions of real variables.

{\it Analogies for the Grad-Shafranov Equation}: When studying the
GS equation, one is confronted with two possible analogies. One is
well known and has been often used in previous works: the analogy
with the Helmholtz problem of mathematical physics. The other is
new and will be presented here for the first time: the analogy
with the CS equation.

Eq. (\ref{GS}) when considered in the traditional way as an
equation for a real function of a real variable $\Psi(x,y)$ is an
application of the Helmholtz problem:
\begin{equation}
\triangle \Upsilon =\lambda \Upsilon \label{realhelmholtz}
\end{equation}
for the eigenvalue $\lambda$ and the eigenfunction $\Upsilon$.

In studies of the GS equation applied to space and astrophysics
systems, the analogy with the Helmholtz problem
(\ref{realhelmholtz}) has been very fruitful in studying arcade
expansion~\cite{lbb}. However, all solutions provided by the
analogy with the Helmholtz problem do not include the presence of
reconnection. As noted above, reconnection is believed to be
present as suggetsed by simulation and supported by observation.
To arrive to a model that incorporates the effects of reconneciton
and the presence of plasmoids, we need to abandon the analogy with
the Helmhotlz equation and propose a new one: the analogy with the
CS equation.

The inductive process that leads from the Helmholtz problem to the
CS equation is typical of the discipline of nonlinear propagation
in  optical waveveguides. Wave propagation in optical waveguides
is governed by a Helmholtz equation for the electromagnetic
field~\cite{okamoto}:
\begin{equation} \triangle \Psi =- k^2 n^2
\Psi \label{guideHelmholtz}
\end{equation}
where $n$ is the refractive index of the media, $k$ the wavenumber
 and $\Psi$ is a complex function, interpreted physically as a
polarization component of the electromagnetic field. In non-homogeneous
media $n$ is a function of space and in non-linear
optical materials it is a function of the electric field
amplitude: $n=n(|\Psi|)$. In the most common case of the Kerr
effect~\cite{okamoto}, the dependence of the refractive index upon
the amplitude of the field is quadratic: $n=n_0 + n_2 |\Psi|^2$.
Following the classic textbook derivation~\cite{okamoto}, we
assume a 2D system (in analogy with the system under consideration
for the GS equation) and we rewrite the unknown function using a
leading harmonic dependence of the field along the propagating
direction (that we assume to be $y$ in the present case):
\begin{equation}
\Psi(x,y) = \varphi(x,y) e^{-j k n_0 y}
\end{equation}
where $j$ is the imaginary unit. Absolute generality is still
provided by retaining the most general dependence of the factor
$\varphi(x,y)$. For the new unknown function $\varphi$,  the
complex Helmholtz equation (\ref{guideHelmholtz}) yields:
\begin{equation}
\frac{\partial^2 \varphi}{\partial x^2}+\frac{\partial^2
\varphi}{\partial y^2} -2 j k n_0 \frac{\partial \varphi}{\partial
y} +k^2 (n^2-n_0^2) \varphi =0 \label{guide}
\end{equation}
Traditionally, a further approximation is made~\cite{okamoto},  assuming
$(n^2-n_0^2)\cong 2n_0(n-n_0)$ and requiring the applicability of
the paraxial (or Fresnel) approximation that prescribes
$|\frac{\partial^2 \varphi}{\partial y^2}|\ll 2 k n_0
|\frac{\partial \varphi}{\partial y}| $. The final equation
typically used to study light propagation in non linear media
becomes:
\begin{equation}
 \frac{\partial \varphi}{\partial y}= -j \frac{1}{2kn_0}\frac{\partial^2
\varphi}{\partial x^2}  -j  k n_2 |\varphi|^2 \varphi \label{CS}
\end{equation}
none others than the cubic Schr\"odinger (CS)
equation~\cite{lamb}. A well known theoretical result is that the
CS equation above admits a soliton solution of the
form~\cite{lamb,okamoto}:
\begin{equation}
\varphi =\varphi_P {\rm sech}\left(\frac{x}{y_0}\right) \exp
\left(\frac{-jy}{2n_0 y_0^2}\right) \label{solitonCS}
\end{equation}
where $y_0$ is a free parameter and $|\varphi_P|^2= 2/n_0 n_2
y_0^2$.

The  formal analogy of the  original Helmholtz equation in the
complex plane used to study light propagation with the GS equation
allows to derive a new class of soliton-like solution similar to
eq.~(ref{solitonCS}).

{\it Soliton solutions of the  Grad-Shafranov Equation}: To use
the analogy with the CS equation outlined above, we consider the
GS equation in its original form (\ref{GSdue}) and assume the
following choice of the free functions $B_z$ and $p$:
\begin{equation}
\begin{array}{l}
\displaystyle B_z \nabla B_z= \alpha_0^2 \Psi \nabla \Psi \\ \\
\displaystyle  \nabla p= \alpha_0^2 |\Psi|^2\Psi \nabla \Psi
\end{array} \label{choice}
\end{equation}
The GS equation  (\ref{GSdue}) is extended in the complex plane
and complex functions $\Psi$ of real variables $(x,y)$ are sought
as solutions. The choice above (\ref{choice}) is a possible and
legitimate choice of the free functions in the GS equation.
Although at first sight it might appear arbitrary, the choice
selected above has a simple physical meaning: it ensures that
$\nabla B^2_z/\nabla p$ is independent of the $y$ coordinate along
which the soliton-like solution propagates. With this choice the
first of the GS eq.~(\ref{GSdue}) is automatically satisfied and
the second becomes identical to eq.(\ref{guideHelmholtz}) with a
non-linear refractive index of the form prescribed by the Kerr
effect:
\begin{equation}
\frac{\partial^2 \Psi}{\partial x^2}+\frac{\partial^2
\Psi}{\partial y^2}  = -\alpha_0^2 (1+ |\Psi|^2) \varphi
\label{GScomplex}
\end{equation}
A soliton solution can be readily obtained in the form:
\begin{equation}
\Psi(x,y)=\Psi_p {\rm sech}(x/L)e^{-j(\alpha_0+1/2\alpha_0y_0^2)y}
\label{GSsoliton}
\end{equation}
with $L=(1/y_0^2+1/4\alpha_0^2y_0^4)^{-1/2}$ and
$\Psi_p=\sqrt{2}/\alpha_0L$.

Note that the soliton solution (\ref{GSsoliton}) is exact and no
paraxial approximation has been made in the present case, since
the exact solution can be found even for the exact GS equation and
the physical significance of the paraxial approximation is
peculiar to optical waveguide propagation, and of no meaning in
the case of the GS equation.

\begin{figure}
\centering
\includegraphics[width=60mm,angle=0]{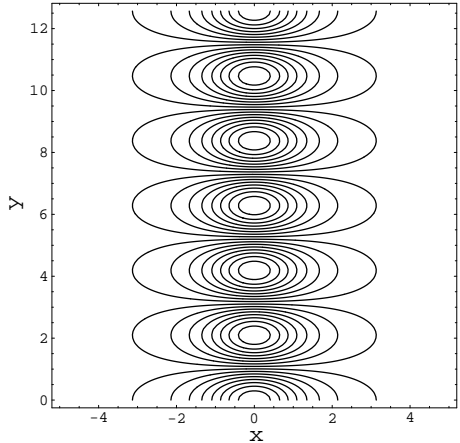}
\caption{Contour plot of the real part of the flux function,
corresponding to the section of the magnetic surfaces on the
$(x,y)$ plane.
 } \label{fieldlines}
\end{figure}

Figure~\ref{fieldlines} shows the contour lines of the real part
of the flux function $\Psi$, physically such contour lines
correspond to the section of the magnetic surfaces on the
$(x,y)$plane. The field lines belong on the flux surfaces but have
a third component away from the plane and given by $B_z$:
\begin{equation}
B_z=\alpha_0 \Psi \label{bz}
\end{equation}

The magnetic field topology presents a series of magnetic islands
piled one above the other. Physically, this solution can be
considered the end result of a repetitive series of processes of
arcade stretching and plasmoid ejections that compose a jet of
plasma and magnetic field emitted from a differentially rotating
surface (such as the photosphere or an accretion disk as described
above).

It is interesting to observe that the primary balance of force in
the present case is between the magnetic tension of the field
lines and the plasma pressure. Using the solution
(\ref{GSsoliton}), the initial choice (\ref{choice}) for the free
functions in the GS equation yield an equation for the pressure
perturbation:
\begin{equation}
\triangle \delta p=\nabla \cdot \left(\alpha_0^2 |\Psi|^2 \Psi
\nabla \Psi \right) \label{pressure}
\end{equation}
In astrophysical systems, it is reasonable to assume a background
pressure that is being perturbed by the magnetic
structure~\cite{lyndenbell}. Eq.~(\ref{pressure}) can be readily
solved observing that the $y$ dependence of $\delta p$ is
harmonic:
\begin{equation}
\delta p=\chi(x) e^{-2j(\alpha_0+1/2\alpha_0y_0^2)y}
\label{pharmon}
\end{equation}
Substituting into eq.~(\ref{pressure}), a simple ordinary equation
is obtained for the $x$-dependent factor $\chi$ of the pressure
perturbation:
\begin{equation}
\begin{array}{r}
\displaystyle \frac{d^2\chi}{dx^2}-4 \left(\alpha_0+\frac{1}{2
\alpha_0 y_0^3}\right)^2 \chi= -\left(\frac{{\rm sech}^3(x/L)}{4
\alpha_0^4 y_0^3L^2}\right)^2  \cdot \\ \\
\displaystyle \left( 4 \left(\alpha_0^4 y_0^2 +
\frac{1}{L^2}\right) + \left(4\alpha_0^4 y_0^2 -
\frac{1}{L^2}\right) \cosh(2x/L)\right)
\end{array}
\label{pressurex}
\end{equation}
that is solved numerically.

\begin{figure}
\centering
\includegraphics[width=60mm,angle=0]{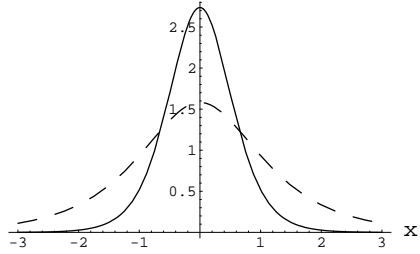}
\caption{Dependence on $x$ of the pressure  (solid) and of the flux
function (dashed), $\alpha_0=1$, $y_0=1$. } \label{chi}
\end{figure}

Figure~\ref{chi} shows the $x$ dependence of the pressure $\chi$.
The $y$ dependence is harmonic and is given by eq.~(\ref{pharmon})

{\it Existence in Nature of soliton-like solutions}: The solution
derived above is an exact and legitimate solution of the GS
equation. It has been derived in the complex plane and to obtain a
physically viable solution one has simply to take either the real
or the imaginary part.

Once we have shown the mathematical existence of soliton-like
solutions, we need to consider the possibility that in practice
they can be identified as realistic descriptions of processes
observed in nature or generated in laboratory. Here we propose
that soliton-like solutions can explain the observed structure of
astrophysics jets. Astrophysics jets are observed to be emitted
from rotating accretion disks. The dynamo effect inside the
accretion disks generate a magnetic field that is then emitted
with the flow of the jet~\cite{lyndenbell}. Observations show that
often jets present a so called "knotty" structure where the jet is
composed by a number of islands (knots)~\cite{knot}. This feature
suggests the possibility to model the magnetic structure of jets
using multiple islands solutions of the Grad-Shafranov equation.
While the solution-like solutions are by no means the only
solutions presenting multiple islands (e.g. ~\cite{bogo} for
another example), they present the right qualitative behaviour
observed in astrophysics jets. Two issues need to be considered to
propose credibly the applicability of soliton-like solutions.

The first question is whether a soliton-like solution is stable.
Previous studies have considered the stability of equilibria
composed by periodic magnetic islands, proposing the existence
of the coalescence instability~\cite{pritchett}. For sufficiently
large aspect ratio islands, i.e. for islands stretched along the
axis of the jet (the $y$ axis of the solution above) the
instability is greatly reduced allowing the soliton-like solution
to remain stable for long times (that observationally turn into
long distances, recalling that the jets are moving at relativistic
speeds). One can indeed propose that the eventual onset of this
instability can explain why the jets after remaining collimated
for kiloparsec distances eventually destabilize in large magnetic
clouds.

The second question is whether a soliton-like solution can be
created by a disk. To answer this question one should do a
simulation of the accretion disk and of the jets emitted from
them. A complete treatment is extremely challenging and
beyond the scope of the present work (if at
all possible). Yet limited studies that focus on just a part of
the problem are possible and have been performed. The creation of
repetitive islands similar to the islands featured by the
soliton-like solution have indeed been observed in simulations of
a problem closely related to jet formation: the study of plasma
emitted from the solar corona~\cite{choe}. And indeed the ejection
of multiple magnetic islands has been observed for the jet from
SS433~\cite{chakrabarti}


\begin{thebibliography}{99}


\bibitem{biskamp}  D. Biskamp, {\it Nonlinear magnetohydrodynamics } (Cambridge University Press, Cambridge, 1993).


\bibitem{priest}  E.R. Priest, T. Forbes, {\it Magnetic Reconnection: MHD Theory and Applications} (Cambridge University Press, Cambridge, 1999).

\bibitem{lyndenbell}  D. Lynden-Bell, {\it Mon. Not. R. Astron. Soc.}, {\bf 279}, 389 (1996).

\bibitem{lbb}  D. Lynden-Bell,C. Boily,  {\it Mon. Not. R. Astron. Soc.}, {\bf 267}, 146 (1994).

\bibitem{choe}  G.S. Choe, C.Z. Cheng,  {\it Astrphys. J.}, {\bf 541}, 449 (2000).

\bibitem{chakrabarti}  S.K. Chakrabarti, P. Goldoni, P.J. Wiita, A. Nandi, S. Das, {\it Astrphys. J.}, {\bf 576}, L45 (2002).

\bibitem{knot}  A.H. Bridle, D.H. Hough, C.J. Lonsdale, J.O.
Burns, R.A. Laing, {\it Astron. J.}, {\bf 108}, 766 (1994).

\bibitem{lamb}  G.L. Lamb, {\it Elements of Soliton Theory} (Wiley, New York, 1980).


\bibitem{finn}  J.M. Finn, W.M. Manheimer, E. Ott,  {\it Phys. Fluids}, {\bf 24}, 1336 (1981).

\bibitem{okamoto}  K. Okamoto, {\it Fundamentals of Optical Waveguides} (Academic Press, San Diego, 2000).


\bibitem{bogo} O.I. Bogoyavlenskij, {\it Phys. Rev. Lett.}, {\bf 84}, 1914 (2000).

\bibitem{pritchett}  P.L. Pritchett, C.C. Wu,  {\it Phys. Fluids}, {\bf 22}, 2140 (1979).

\end{thebibliography}
\end{document}